\documentclass[prb,showpacs,floatfix,twocolumn,byrevtex,superscriptaddress]{revtex4-1}

\usepackage[english]{babel}
\usepackage{amsmath}
\usepackage{graphicx}
\usepackage{float}
\usepackage{bm}
\usepackage{multirow}
\usepackage{dcolumn}

\newcommand{\icm}{\ensuremath{~\textrm{cm}^{-1}}}% % cm-1
\newcommand{\BKFA}{Ba$_{0.6}$K$_{0.4}$Fe$_{2}$As$_{2}$}
\newcommand{\BKFAx}{Ba$_{1-x}$K$_{x}$Fe$_{2}$As$_{2}$}
\newcommand{\BFA}{BaFe$_{2}$As$_{2}$}

\begin{document}

\title{Pseudogap in underdoped Ba$_{1-x}$K$_{x}$Fe$_{2}$As$_{2}$ as seen via optical conductivity}
\author{Y. M. Dai}
\affiliation{LPEM, ESPCI-ParisTech, CNRS, UPMC, 10 rue Vauquelin, F-75231 Paris Cedex 5, France}
\affiliation{Beijing National Laboratory for Condensed Matter Physics, National Laboratory for Superconductivity, Institute of Physics, Chinese Academy of Sciences, P.O. Box 603, Beijing 100190, China}
\author{B. Xu}
\affiliation{LPEM, ESPCI-ParisTech, CNRS, UPMC, 10 rue Vauquelin, F-75231 Paris Cedex 5, France}
\affiliation{Beijing National Laboratory for Condensed Matter Physics, National Laboratory for Superconductivity, Institute of Physics, Chinese Academy of Sciences, P.O. Box 603, Beijing 100190, China}
\author{B. Shen}
\affiliation{Beijing National Laboratory for Condensed Matter Physics, National Laboratory for Superconductivity, Institute of Physics, Chinese Academy of Sciences, P.O. Box 603, Beijing 100190, China}
\author{H. H. Wen}
\affiliation{Beijing National Laboratory for Condensed Matter Physics, National Laboratory for Superconductivity, Institute of Physics, Chinese Academy of Sciences, P.O. Box 603, Beijing 100190, China}
\affiliation{National Laboratory of Solid State Microstructures and Department of Physics, Nanjing University, Nanjing 210093, China}
\author{J. P. Hu}
\affiliation{Beijing National Laboratory for Condensed Matter Physics, National Laboratory for Superconductivity, Institute of Physics, Chinese Academy of Sciences, P.O. Box 603, Beijing 100190, China}
\affiliation{Department of Physics, Purdue University, West Lafayette, Indiana 47907, USA}
\author{X. G. Qiu}
\email[]{xgqiu@aphy.iphy.ac.cn}
\affiliation{Beijing National Laboratory for Condensed Matter Physics, National Laboratory for Superconductivity, Institute of Physics, Chinese Academy of Sciences, P.O. Box 603, Beijing 100190, China}
\author{R. P. S. M. Lobo}
\email[]{lobo@espci.fr}
\affiliation{LPEM, ESPCI-ParisTech, CNRS, UPMC, 10 rue Vauquelin, F-75231 Paris Cedex 5, France}

\date{\today}
\begin{abstract}
We report the observation of a pseudogap in the \emph{ab}-plane optical conductivity of underdoped \BKFAx\ ($x = 0.2$ and 0.12) single crystals. Both samples show prominent gaps opened by a spin density wave (SDW) order and superconductivity at the transition temperatures $T_{\it SDW}$ and  $T_c$, respectively. In addition, we observe an evident pseudogap below $T^{\ast} \sim$ 75 K, a temperature much lower than $T_{\it SDW}$ but much higher than $T_{c}$. A spectral weight analysis shows that SDW competes with superconductivity while the pseudogap is closely connected to the superconducting gap, indicating the possibility of its being a precursor of superconductivity. The doping dependence of the gaps is also supportive of such a scenario.
\end{abstract}
\pacs{74.25.Gz, 74.70.Xa, 78.30.-j}
\maketitle

%%%%%%%%%%%%%%%%%%%%%%%%%%%%%%%%%%%%%%%%%%%%%%%%%%%%%%%%%%%%%%%%%%%%%%%%%%%%%%%
%
% Introduction
%

Among all the families of iron-pnictide superconductors discovered to date, \cite{Kamihara2008,Torikachvili2008,Rotter2008,Sefat2008,Li2009,Tapp2008,Hsu2008} the \BFA\ (Ba122) family is one of the most studied. The parent \BFA\ composition is a poor Pauli-paramagnetic metal with a structural and magnetic phase transition at 140 K. \cite{Rotter2008a} Superconductivity arises with the suppression of magnetism that can be achieved by applying pressure \cite{Torikachvili2008} or chemical substitution. \cite{Rotter2008,Sefat2008,Li2009} The substitution of Ba with K atoms yields hole-doping \cite{Rotter2008} with a maximum $T_{c} \approx 39$~K and  the substitution of Fe atoms by Co or Ni atoms results in electron-doping \cite{Sefat2008,Li2009} with a maximum $T_{c} \approx 25$~K. Extensive studies have been carried out in the parent \BFA, \cite{Hu2008,Akrap2009} electron-doped BaFe$_{2-x}A_x$As$_2$ ($A$ = Co, Ni), \cite{Lobo2010,Tu2010,Teague2011,Terashima2009} as well as optimally hole-doped \BKFA\ \cite{Li2008,Charnukha2011a,Dai2011,Shan2011,Ding2008} compounds. However, the hole-underdoped regime of the phase diagram is relatively unexplored.

This hole-underdoped region is of the utmost importance. Firstly, the superconducting mechanism is deeply tied with magnetism. The interplay between magnetism and superconductivity is manifest in this regime, where SDW and superconductivity coexist over a large doping range. \cite{Park2009,Goko2009,Aczel2008,Massee2009,Chia2010} Secondly, in cuprates, the most exciting, yet puzzling, physics takes place in the hole-underdoped regime. This regime thus is pivotal to the comparison between iron-pnictides and cuprates.

Xu \emph{et al.} angle-resolved photoemission (ARPES) measurements on underdoped \BKFAx\ \cite{Xu2011} showed a distinct pseudogap coexisting with the superconducting gap and suggested that both the pseudogap and superconductivity are driven by antiferromagnetic fluctuations. However, one key issue in understanding the origin of the pseudogap and, in particular, its relation to superconductivity is the question of whether it shares electronic states with the superconducting condensation. \cite{Yu2008} Infrared spectroscopy probes the charge dynamics of bulk materials and the spectral weight analysis is a powerful tool to address this issue.

We measured broadband infrared spectroscopy measurements on two underdoped \BKFAx\ ($x = 0.2$ and 0.12) single crystals. In both samples, the opening of the SDW gap and the superconducting gap was clearly observed on the optical conductivity. In addition, another small gap opens below $T^{\ast} \sim$ 75 K, closely resembling the pseudogap in the hole-underdoped cuprates. We find that the SDW gap depletes the spectral weight available for the superconducting condensate, which suggests that the SDW order competes with superconductivity. On the other hand, the pseudogap shares electronic states with the superconducting gap. This is shown by the doping and temperature dependence of the optical conductivity spectral weight. This observation supports a scenario where the pseudogap is a precursor of superconductivity.

%%%%%%%%%%%%%%%%%%%%%%%%%%%%%%%%%%%%%%%%%%%%%%%%%%%%%%%%%%%%%%%%%%%%%%%%%%%%%%%
%
% Methods
%

High quality \BKFAx\ single crystals were grown by the self-flux method using FeAs as the flux. \cite{Luo2008}
\begin{figure}[htb]
  \includegraphics[width=\columnwidth]{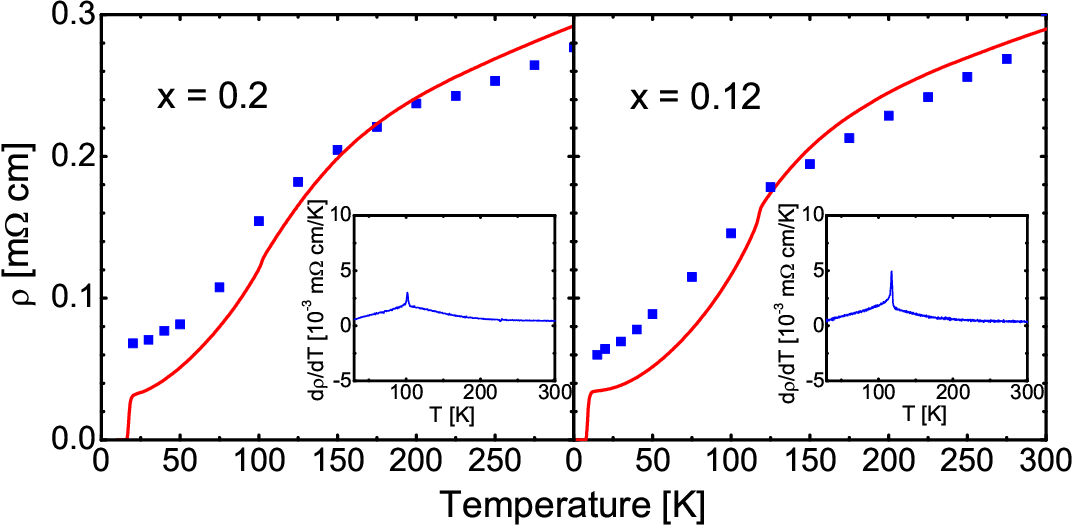}
  \caption{ (color online) Left panel: Temperature dependence of the resistivity of \BKFAx\ ($x = 0.2$) single crystal (red solid line). A steep superconducting transition can be seen at $T_c = 19$~K. The blue solid squares are values from the zero frequency extrapolation of the optical conductivity. The inset shows the derivative of the resistivity $d\rho/dT$ as a function of temperature. The sharp peak at 104 K in $d\rho/dT$ is associated with the SDW transition. The right panel depicts the same curves for $x = 0.12$ sample with $T_c = 11$~K, and $T_{\it SDW} = 121$~K.
}
  \label{RT}
\end{figure}
The left panel of Fig.~\ref{RT} shows the temperature dependence of the DC resistivity [$\rho(T)$] for the \BKFAx\ ($x = 0.2$) sample. The $\rho(T)$ curve is characterized by a steep superconducting transition at $T_{c}$ = 19 K. The inset shows the derivative of the resistivity $d\rho / dT$ as a function of temperature. The SDW transition manifests itself as a sharp peak in $d\rho / dT$ at $T_{\it SDW} = 104$~K, which corresponds to a small kink on the $\rho(T)$ curve. The right panel displays the same curves for the $x = 0.12$ sample, which has $T_{c} = 11$~K, and $T_{\it SDW} = 121$~K.

The \emph{ab}-plane reflectivity [$R(\omega)$] was measured at a near-normal angle of incidence on Bruker IFS113v and IFS66v spectrometers. An \emph{in situ} gold overfilling technique \cite{Homes1993} was used to obtain the absolute reflectivity of the samples. Data from 20 to 12000\icm\ were collected at 18 different temperatures from 5 to 300 K on freshly cleaved surfaces. In order to use Kramers-Kronig analysis, we extended the data to the visible and UV range (10000 to 55000\icm) at room temperature with an AvaSpec-2048 $\times$ 14 model fiber optic spectrometer.

%%%%%%%%%%%%%%%%%%%%%%%%%%%%%%%%%%%%%%%%%%%%%%%%%%%%%%%%%%%%%%%%%%%%%%%%%%%%%%%
%
% Results
%

\begin{figure}[htb]
  \includegraphics[width=0.8\columnwidth]{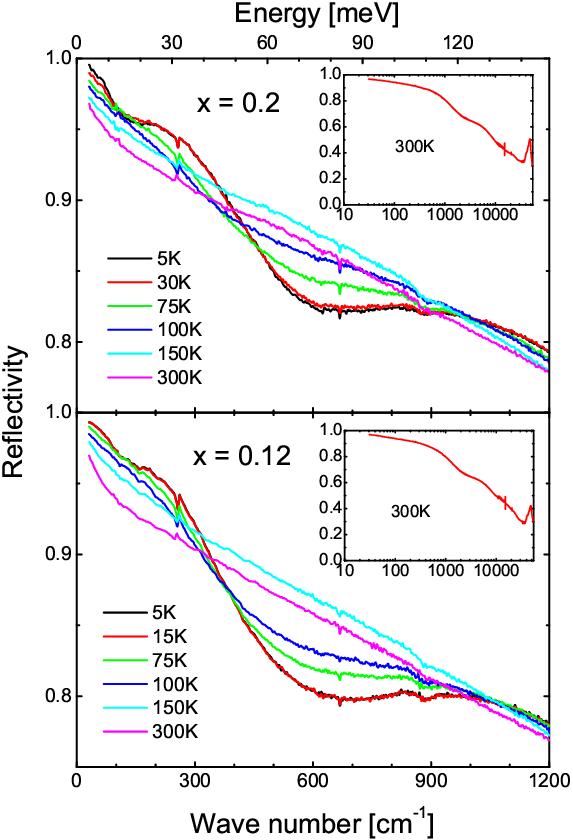}
  \caption{ (color online) Reflectivity of \BKFAx\ single crystals below 1200\icm\ at various temperatures. Top panel: $x = 0.2$; Bottom panel: $x = 0.12$. Inset: Reflectivity of full measured range at 300~K.}
  \label{Reflec}
\end{figure}

Figure \ref{Reflec} shows the infrared reflectivity at selected temperatures for both samples up to 1200\icm. The inset in each panel displays the reflectivity for the full measured range at 300 K. For the $x = 0.2$ sample, shown in the top panel, the reflectivity exhibits a metallic response and approaches unity at zero frequency. Below $T_{\it SDW} = 104$~K, a substantial suppression of $R(\omega)$ at about 650\icm\ sets in and intensifies with decreasing temperature.
Simultaneously, the low frequency reflectivity continues increasing towards unity. This is a signature of a partial SDW gap on the Fermi surface. Below 75~K, defined as $T^{\ast}$ here, another suppression of $R(\omega)$ appears in a lower energy scale ($\sim 150\icm$) signaling the opening of a second partial gap (pseudogap) with a smaller value. Upon crossing the superconducting transition, the reflectivity below $\sim 150\icm$ increases indicating the opening of a superconducting gap. Similar features are observed on $R(\omega)$ for the $x = 0.12$ sample as shown in the bottom panel of Fig.~\ref{Reflec}.

The real part of the optical conductivity $\sigma_{1}(\omega)$ was determined by Kramers-Kronig analysis of the measured reflectivity.
\begin{figure}[tb]
  \includegraphics[width=0.8\columnwidth]{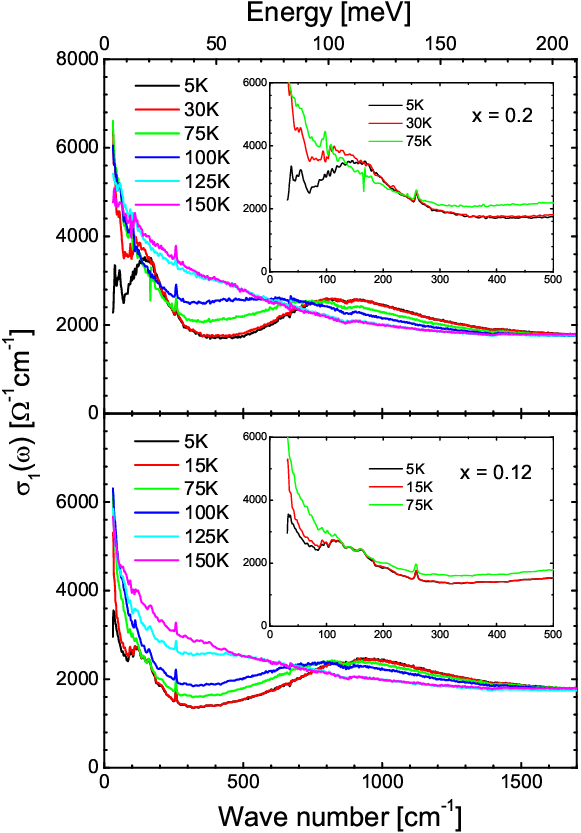}
  \caption{ (color online) Top panel: Optical conductivity of \BKFAx\ ($x = 0.2$) at selected temperatures. The inset shows the enlarged view of the optical conductivity at low frequencies. The bottom panel displays the same spectra for the $x = 0.12$ sample.
}
  \label{Sigma1}
\end{figure}
Figure \ref{Sigma1} shows $\sigma_{1}(\omega)$ at different temperatures for the two samples. The zero frequency extrapolations of $\sigma_{1}(\omega)$ represent the inverse dc resistivity of the sample, shown as blue solid squares in Fig.~\ref{RT}, which are in good agreement with the transport measurement. The top panel of Fig.~\ref{Sigma1} shows $\sigma_{1}(\omega)$ for \BKFAx\ ($x = 0.2$) below 1700\icm. At 150 K and 125 K, hence above $T_{\it SDW}$, a Drude-like metallic response dominates the low frequency optical conductivity. Below $T_{\it SDW}$, $\sigma_{1}(\omega)$ below about 650\icm\ is severely suppressed. Meanwhile, it increases in a higher energy scale from 650\icm\ to 1700\icm. The optical conductivity for the normal state and that for the SDW state just below $T_{\it SDW}$ show an intersection point at about 650\icm.  As the temperature decreases, both the low energy spectral suppression and the high energy bulge become stronger, resulting in a prominent peak at about 840\icm. This spectral evolution manifests the behavior of the SDW gap in this material: transfer of low frequency spectral weight to high frequencies. If we take the peak position as an estimate of the gap value, we can see that the gap increases with decreasing temperature. The effect of the SDW transition on optical conductivity has been well established in the parent \BFA, \cite{Hu2008,Nakajima2010,Pfuner2009} which is characterized by two distinct SDW gaps, associated with different Fermi surfaces, at 360 and 890\icm, respectively.

Below $T^{\ast} \sim 75$~K, a second suppression in the optical conductivity sets in below roughly 110\icm\ with a bulge extending from about 110\icm\ to 250\icm, implying the opening of the pseudogap on the Fermi surface. The inset of Fig.~\ref{Sigma1} shows the enlarged view of the low temperature optical conductivity at low frequencies, where the pseudogap is seen more clearly. This pseudogap is unlikely due to the SDW transition as (i) it opens at a temperature between 50 K and 75 K, well below $T_{\it SDW}$ and (ii) it redistributes spectral weight at a different, much lower energy scale, which is about the energy scale of the superconducting gap in the optimally doped \BKFA. \cite{Li2008,Charnukha2011a,Dai2011}

The superconducting transition at $T_{c}$ = 19 K implies the opening of a superconducting gap. As shown in the inset of Fig.~\ref{Sigma1}, this leads to the reduction of the optical conductivity at low frequencies between 30 K and 5 K. The spectral weight lost in the transition is recovered by the $\delta(\omega)$ function at zero frequency representing the infinite DC conductivity in the superconducting state. This  $\delta(\omega)$ function is not visible in the $\sigma_{1}(\omega)$ spectra, because only finite frequencies are experimentally measured. Nevertheless, its weight can be calculated from the imaginary part of the optical conductivity. \cite{Zimmers2004,Dordevic2002} Note that, the spectral depletion in $\sigma_1(\omega)$ due to the superconducting condensate extends up to 180\icm. This is the same energy scale of the pseudogap, hinting that the superconducting gap and the pseudogap share the same electronic states, and may have the same origin.

It is also noteworthy that, in the superconducting state, large residual low frequency conductivity, representing unpaired quasiparticles, is present, even at the lowest measured temperature 5 K. This is in sharp contrast to the optimally doped \BKFA, where the low frequency optical conductivity is dominated by a fully open gap. \cite{Li2008,Charnukha2011a,Dai2011} Here, the residual low frequency conductivity observed in the underdoped \BKFAx\ might be due to the well-known phase separation \cite{Park2009,Goko2009,Julien2009} or nodes in the superconducting gap. \cite{Reid2011,Maiti2012}

In the $x = 0.12$ sample, very similar features are observed, as shown in the bottom panel of Fig.~\ref{Sigma1}, but remarkable differences exist: (i) the SDW gap opens at a higher temperature ($T_{\it SDW} = 121$~K) and the gap value shifts to a higher energy scale ($\sim 890\icm$); (ii) The low frequency spectral weight suppression due to the SDW gap is stronger, indicating that a larger part of the Fermi surface is removed in the SDW state; (iii) In contrast to the SDW gap, both the pseudogap and the superconducting gap features are weaker in the more underdoped sample. The evolution of the three gaps (SDW, pseudogap and superconducting) with doping also suggests that the pseudogap and the superconducting gap may have a common origin while the SDW is a competitive order to superconductivity.

In order to investigate the origin of the pseudogap and the relationship among all gaps, we analyzed the data utilizing a restricted spectral weight, defined as:
\begin{equation}
\label{EqRSW}
SW_{\omega_a}^{\omega_b} = \int_{\omega_a}^{\omega_b} \sigma_{1}(\omega) d \omega,
\end{equation}
where $\omega_a$ and $\omega_b$ are lower and upper cut-off frequencies, respectively. By choosing appropriate values for $\omega_a$ and $\omega_b$, one can study the relations among different phase transitions. \cite{Yu2008} When replacing $\omega_a$ by 0 and $\omega_b$ by $\infty$, we fall back to the standard $f$-sum rule and the spectral weight is conserved.

\begin{figure}[tb]
  \includegraphics[width=0.8\columnwidth]{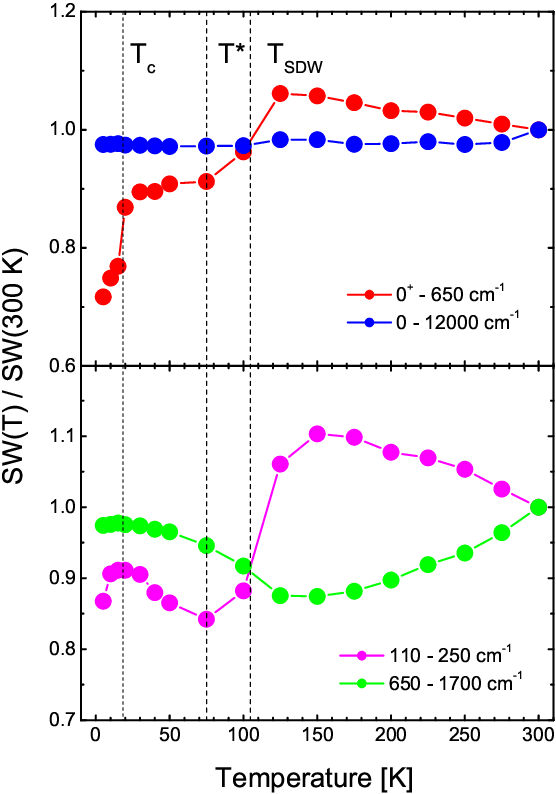}
  \caption{ (color online) Temperature dependence of the spectral weight, SW$_{\omega_a}^{\omega_b}$ = $\int_{\omega_a}^{\omega_b}\sigma_{1}(\omega)d\omega$, between different lower and upper cutoff frequencies for the $x = 0.2$ sample. The vertical dashed lines denote $T_{c}$, $T^{\ast}$ and $T_{\it SDW}$.}
  \label{RSW}
\end{figure}
Figure \ref{RSW} shows the temperature dependence of the $x = 0.2$ sample spectral weight, normalized by its value at 300 K, at different cut-off frequencies. The vertical dashed lines denote $T_{c}$, $T^\ast$ and $T_{\it SDW}$. The blue solid circles in the top panel of Fig.~\ref{RSW} are the normalized spectral weight with cut-off frequencies $\omega_a = 0$ and $\omega_b = 12000\icm$ as a function of temperature. Here, the weight of the zero frequency $\delta$-function is included below $T_{c}$. Moreover, since the optical conductivity is measured only down to 20\icm, we estimate the spectral weight below that energy by fitting the low frequency normal state optical conductivity to a Drude model. The upper cut-off frequency ($\omega_b = 12000\icm$) is high enough to cover the whole spectrum responsible for the phase transitions in this material. Hence the blue solid circles form a flat line at about unity, indicating that the spectral weight is conserved.

The red solid circles in the top panel show the temperature dependence of the normalized spectral weight with cut-off frequencies $\omega_a = 0^{+}$ and $\omega_b = 650\icm$. Here $0^{+}$ means that the superfluid weight is not included. Above $T_{\it SDW}$, the continuous increase of the normalized SW$_{0^{+}}^{650}$ with decreasing $T$ is related to the narrowing of the Drude band. This is the typical optical response of a metallic material. A strong spectral weight suppression occurs at $T_{\it SDW}$, which is the consequence of the SDW gap opening. At $T_{c}$, another sharp drop of the spectral weight breaks in, indicating the superconducting gap opening.

The temperature dependence of the normalized SW$_{650}^{1700}$, shown as green solid circles in the bottom panel of Fig.~\ref{RSW}, provides clues about the relation between the superconducting and the SDW gaps. Above $T_{\it SDW}$, the material shows a metallic response which can be described by a Drude peak centered at zero frequency. With the temperature decrease, the DC conductivity increases and the scattering rate reduces. The continuous narrowing of the Drude band induces a transfer of spectral weight from the mid-infrared to the far infrared, resulting in the continuous decrease of the spectral weight observed in the 650--1700\icm\ range. Below $T_{\it SDW}$, the opposite behavior dominates the optical conductivity. The SDW gap depletes the spectral weight below 650\icm\ and transfers it to the 650--1700\icm\ range, leading to the continuous increase of SW$_{650}^{1700}$ with decreasing $T$. This behavior continues into the superconducting state and does not show any feature at $T_{c}$. These observations indicate that the SDW and superconducting gaps are separate and even act as competitive orders in this material.

If a partial gap is due to a precursor order of superconductivity, for example preformed pairs without phase coherence, once the long range superconductivity is established, a significant part of the spectral weight transferred to high frequencies by the partial gap should be transferred back to low energies and join the superconducting condensate. \cite{Ioffe1999,Ioffe2000} Conversely, a partial gap due to a competitive order to superconductivity depletes the low-energy spectral weight and holds it in a high energy scale without transferring it back to the superfluid weight below $T_{c}$. \cite{Yu2008} From the normalized SW$_{650}^{1700}$ \emph{vs} $T$ curve (green solid circles) we note that no loss of spectral weight is observed at $T_{c}$. This means that the spectral weight transferred to high frequencies by the SDW gap remains in the high frequency scale and does not contribute to the superconducting condensate. Therefore, the SDW acts as a competitive order to superconductivity in this material.

Along these lines, the origin of the pseudogap and its relationship to superconductivity can be revealed by a close inspection of the temperature dependence of the normalized SW$_{110}^{250}$, shown as pink solid circles in the bottom panel. Above $T^{*}$, this curve shows the same feature as the normalized SW$_{0^{+}}^{650}$ \emph{vs} $T$ curve, \emph{i.e.}, continuous increase upon cooling down followed by a suppression at $T_{\it SDW}$ due to the SDW gap opening. At $T^{*}$, the spectral weight in the 110--250\icm\ range reaches a minimum and starts to increase with decreasing temperature. This is due to the opening of the pseudogap. The pseudogap, opening at $T^{*}$, depletes the spectral weight below 110\icm\ and retrieves it in the 110--250\icm\ frequency range, leading to the increase of SW$_{110}^{250}$ below $T^{*}$. An interesting phenomenom happens to the pseudogap when the material undergoes the superconducting transition. In contrast to the case of the SDW gap, a significant loss of spectral weight in the 110--250\icm\ frequency range is observed below $T_{c}$. This observation indicates that the spectral weight transferred to the 110--250\icm\ range by the pseudogap joins the superconducting condensate when superconductivity is established. Hence, the pseudogap is likely a precursor order with respect to superconductivity.

A recent ultrafast pump-probe study of underdoped \BKFAx\ revealed a normal-state precursor order to superconductivity at $T^{\ast} \sim$ 60 K. \cite{Chia2010} The precursor scenario is also supported by $^{75}$As NQR measurements on Ca(Fe$_{1-x}$Co$_{x}$)$_{2}$As$_{2}$ showing a pseudogap-like phase as a precursor state for the coherent superconducting phase \cite{Baek2011} and Andreev reflection studies in Ba(Fe$_{1-x}$Co$_{x}$)$_{2}$As$_{2}$ which offered evidence for phase-incoherent superconducting pairs in the normal state. \cite{Sheet2010}

%%%%%%%%%%%%%%%%%%%%%%%%%%%%%%%%%%%%%%%%%%%%%%%%%%%%%%%%%%%%%%%%%%%%%%%%%%%%%%%
%
% Conclusions
%

In summary, we measured the optical conductivity of two underdoped \BKFAx\ ($x = 0.2$ and 0.12) single crystals. In both samples, besides the SDW gap and superconducting gap, the optical conductivity reveals another small partial gap (pseudogap) opening below $T^{\ast} \sim$ 75 K an intermediate temperature between $T_{SDW}$ and $T_{c}$. A spectral weight analysis shows that the SDW gap diminishes the low energy spectral weight available for the superconducting condensate while the pseudogap shares the same electronic states with the superconducting gap. These observations, together with the doping dependence of these gaps, suggest the SDW as a competitive order and the pseudogap as a precursor to superconductivity.

%%%%%%%%%%%%%%%%%%%%%%%%%%%%%%%%%%%%%%%%%%%%%%%%%%%%%%%%%%%%%%%%%%%%%%%%%%%%%%%
%
% Acknowledgment
%

We thank Hong Xiao, Lei Shan, Cong Ren, Li Yu and Zhiguo Chen for helpful discussion. Work in Paris was supported by the ANR under Grant No. BLAN07-1-183876 GAPSUPRA. Work in Beijing was supported by the National Science Foundation of China (No. 91121004) and the Ministry of Science and Technology of China (973 Projects No. 2011CBA00107, No. 2012CB821400 and No. 2009CB929102). We acknowledge the financial support from the Science and Technology Service of the French Embassy in China.

%%%%%%%%%%%%%%%%%%%%%%%%%%%%%%%%%%%%%%%%%%%%%%%%%%%%%%%%%%%%%%%%%%%%%%%%%%%%%%%%%
%
% The bibliography (BibTeX)
%


\begin{thebibliography}{39}
\expandafter\ifx\csname natexlab\endcsname\relax\def\natexlab#1{#1}\fi
\expandafter\ifx\csname bibnamefont\endcsname\relax
  \def\bibnamefont#1{#1}\fi
\expandafter\ifx\csname bibfnamefont\endcsname\relax
  \def\bibfnamefont#1{#1}\fi
\expandafter\ifx\csname citenamefont\endcsname\relax
  \def\citenamefont#1{#1}\fi
\expandafter\ifx\csname url\endcsname\relax
  \def\url#1{\texttt{#1}}\fi
\expandafter\ifx\csname urlprefix\endcsname\relax\def\urlprefix{URL }\fi
\providecommand{\bibinfo}[2]{#2}
\providecommand{\eprint}[2][]{\url{#2}}

\bibitem[{\citenamefont{Kamihara et~al.}(2008)\citenamefont{Kamihara, Watanabe,
  Hirano, and Hosono}}]{Kamihara2008}
\bibinfo{author}{\bibfnamefont{Y.}~\bibnamefont{Kamihara}},
  \bibinfo{author}{\bibfnamefont{T.}~\bibnamefont{Watanabe}},
  \bibinfo{author}{\bibfnamefont{M.}~\bibnamefont{Hirano}}, \bibnamefont{and}
  \bibinfo{author}{\bibfnamefont{H.}~\bibnamefont{Hosono}},
  \bibinfo{journal}{J.Am.Chem.Soc.} \textbf{\bibinfo{volume}{130}},
  \bibinfo{pages}{3296} (\bibinfo{year}{2008}).

\bibitem[{\citenamefont{Torikachvili et~al.}(2008)\citenamefont{Torikachvili,
  Bud'ko, Ni, and Canfield}}]{Torikachvili2008}
\bibinfo{author}{\bibfnamefont{M.~S.} \bibnamefont{Torikachvili}},
  \bibinfo{author}{\bibfnamefont{S.~L.} \bibnamefont{Bud'ko}},
  \bibinfo{author}{\bibfnamefont{N.}~\bibnamefont{Ni}}, \bibnamefont{and}
  \bibinfo{author}{\bibfnamefont{P.~C.} \bibnamefont{Canfield}},
  \bibinfo{journal}{Phys. Rev. Lett.} \textbf{\bibinfo{volume}{101}},
  \bibinfo{pages}{057006} (\bibinfo{year}{2008}).

\bibitem[{\citenamefont{Rotter et~al.}(2008{\natexlab{a}})\citenamefont{Rotter,
  Tegel, and Johrendt}}]{Rotter2008}
\bibinfo{author}{\bibfnamefont{M.}~\bibnamefont{Rotter}},
  \bibinfo{author}{\bibfnamefont{M.}~\bibnamefont{Tegel}}, \bibnamefont{and}
  \bibinfo{author}{\bibfnamefont{D.}~\bibnamefont{Johrendt}},
  \bibinfo{journal}{Phys. Rev. Lett.} \textbf{\bibinfo{volume}{101}},
  \bibinfo{pages}{107006} (\bibinfo{year}{2008}{\natexlab{a}}).

\bibitem[{\citenamefont{Sefat et~al.}(2008)\citenamefont{Sefat, Jin, McGuire,
  Sales, Singh, and Mandrus}}]{Sefat2008}
\bibinfo{author}{\bibfnamefont{A.~S.} \bibnamefont{Sefat}},
  \bibinfo{author}{\bibfnamefont{R.}~\bibnamefont{Jin}},
  \bibinfo{author}{\bibfnamefont{M.~A.} \bibnamefont{McGuire}},
  \bibinfo{author}{\bibfnamefont{B.~C.} \bibnamefont{Sales}},
  \bibinfo{author}{\bibfnamefont{D.~J.} \bibnamefont{Singh}}, \bibnamefont{and}
  \bibinfo{author}{\bibfnamefont{D.}~\bibnamefont{Mandrus}},
  \bibinfo{journal}{Phys. Rev. Lett.} \textbf{\bibinfo{volume}{101}},
  \bibinfo{pages}{117004} (\bibinfo{year}{2008}).

\bibitem[{\citenamefont{Li et~al.}(2009)\citenamefont{Li, Luo, Wang, Chen, Ren,
  Tao, Li, Lin, He, Zhu et~al.}}]{Li2009}
\bibinfo{author}{\bibfnamefont{L.~J.} \bibnamefont{Li}},
  \bibinfo{author}{\bibfnamefont{Y.~K.} \bibnamefont{Luo}},
  \bibinfo{author}{\bibfnamefont{Q.~B.} \bibnamefont{Wang}},
  \bibinfo{author}{\bibfnamefont{H.}~\bibnamefont{Chen}},
  \bibinfo{author}{\bibfnamefont{Z.}~\bibnamefont{Ren}},
  \bibinfo{author}{\bibfnamefont{Q.}~\bibnamefont{Tao}},
  \bibinfo{author}{\bibfnamefont{Y.~K.} \bibnamefont{Li}},
  \bibinfo{author}{\bibfnamefont{X.}~\bibnamefont{Lin}},
  \bibinfo{author}{\bibfnamefont{M.}~\bibnamefont{He}},
  \bibinfo{author}{\bibfnamefont{Z.~W.} \bibnamefont{Zhu}},
  \bibnamefont{et~al.}, \bibinfo{journal}{New Journal of Physics}
  \textbf{\bibinfo{volume}{11}}, \bibinfo{pages}{025008}
  (\bibinfo{year}{2009}).

\bibitem[{\citenamefont{Tapp et~al.}(2008)\citenamefont{Tapp, Tang, Lv, Sasmal,
  Lorenz, Chu, and Guloy}}]{Tapp2008}
\bibinfo{author}{\bibfnamefont{J.~H.} \bibnamefont{Tapp}},
  \bibinfo{author}{\bibfnamefont{Z.}~\bibnamefont{Tang}},
  \bibinfo{author}{\bibfnamefont{B.}~\bibnamefont{Lv}},
  \bibinfo{author}{\bibfnamefont{K.}~\bibnamefont{Sasmal}},
  \bibinfo{author}{\bibfnamefont{B.}~\bibnamefont{Lorenz}},
  \bibinfo{author}{\bibfnamefont{P.~C.~W.} \bibnamefont{Chu}},
  \bibnamefont{and} \bibinfo{author}{\bibfnamefont{A.~M.} \bibnamefont{Guloy}},
  \bibinfo{journal}{Phys. Rev. B} \textbf{\bibinfo{volume}{78}},
  \bibinfo{pages}{060505} (\bibinfo{year}{2008}).

\bibitem[{\citenamefont{Hsu et~al.}(2008)\citenamefont{Hsu, Luo, Yeh, Chen,
  Huang, Wu, Lee, Huang, Chu, Yan et~al.}}]{Hsu2008}
\bibinfo{author}{\bibfnamefont{F.-C.} \bibnamefont{Hsu}},
  \bibinfo{author}{\bibfnamefont{J.-Y.} \bibnamefont{Luo}},
  \bibinfo{author}{\bibfnamefont{K.-W.} \bibnamefont{Yeh}},
  \bibinfo{author}{\bibfnamefont{T.-K.} \bibnamefont{Chen}},
  \bibinfo{author}{\bibfnamefont{T.-W.} \bibnamefont{Huang}},
  \bibinfo{author}{\bibfnamefont{P.~M.} \bibnamefont{Wu}},
  \bibinfo{author}{\bibfnamefont{Y.-C.} \bibnamefont{Lee}},
  \bibinfo{author}{\bibfnamefont{Y.-L.} \bibnamefont{Huang}},
  \bibinfo{author}{\bibfnamefont{Y.-Y.} \bibnamefont{Chu}},
  \bibinfo{author}{\bibfnamefont{D.-C.} \bibnamefont{Yan}},
  \bibnamefont{et~al.}, \bibinfo{journal}{Proc. Natl. Acad. Sci.}
  \textbf{\bibinfo{volume}{105}}, \bibinfo{pages}{14262}
  (\bibinfo{year}{2008}).

\bibitem[{\citenamefont{Rotter et~al.}(2008{\natexlab{b}})\citenamefont{Rotter,
  Tegel, Johrendt, Schellenberg, Hermes, and P\"ottgen}}]{Rotter2008a}
\bibinfo{author}{\bibfnamefont{M.}~\bibnamefont{Rotter}},
  \bibinfo{author}{\bibfnamefont{M.}~\bibnamefont{Tegel}},
  \bibinfo{author}{\bibfnamefont{D.}~\bibnamefont{Johrendt}},
  \bibinfo{author}{\bibfnamefont{I.}~\bibnamefont{Schellenberg}},
  \bibinfo{author}{\bibfnamefont{W.}~\bibnamefont{Hermes}}, \bibnamefont{and}
  \bibinfo{author}{\bibfnamefont{R.}~\bibnamefont{P\"ottgen}},
  \bibinfo{journal}{Phys. Rev. B} \textbf{\bibinfo{volume}{78}},
  \bibinfo{pages}{020503} (\bibinfo{year}{2008}{\natexlab{b}}).

\bibitem[{\citenamefont{Hu et~al.}(2008)\citenamefont{Hu, Dong, Li, Li, Zheng,
  Chen, Luo, and Wang}}]{Hu2008}
\bibinfo{author}{\bibfnamefont{W.~Z.} \bibnamefont{Hu}},
  \bibinfo{author}{\bibfnamefont{J.}~\bibnamefont{Dong}},
  \bibinfo{author}{\bibfnamefont{G.}~\bibnamefont{Li}},
  \bibinfo{author}{\bibfnamefont{Z.}~\bibnamefont{Li}},
  \bibinfo{author}{\bibfnamefont{P.}~\bibnamefont{Zheng}},
  \bibinfo{author}{\bibfnamefont{G.~F.} \bibnamefont{Chen}},
  \bibinfo{author}{\bibfnamefont{J.~L.} \bibnamefont{Luo}}, \bibnamefont{and}
  \bibinfo{author}{\bibfnamefont{N.~L.} \bibnamefont{Wang}},
  \bibinfo{journal}{Phys. Rev. Lett.} \textbf{\bibinfo{volume}{101}},
  \bibinfo{pages}{257005} (\bibinfo{year}{2008}).

\bibitem[{\citenamefont{Akrap et~al.}(2009)\citenamefont{Akrap, Tu, Li, Cao,
  Xu, and Homes}}]{Akrap2009}
\bibinfo{author}{\bibfnamefont{A.}~\bibnamefont{Akrap}},
  \bibinfo{author}{\bibfnamefont{J.~J.} \bibnamefont{Tu}},
  \bibinfo{author}{\bibfnamefont{L.~J.} \bibnamefont{Li}},
  \bibinfo{author}{\bibfnamefont{G.~H.} \bibnamefont{Cao}},
  \bibinfo{author}{\bibfnamefont{Z.~A.} \bibnamefont{Xu}}, \bibnamefont{and}
  \bibinfo{author}{\bibfnamefont{C.~C.} \bibnamefont{Homes}},
  \bibinfo{journal}{Phys. Rev. B} \textbf{\bibinfo{volume}{80}},
  \bibinfo{pages}{180502} (\bibinfo{year}{2009}).

\bibitem[{\citenamefont{Lobo et~al.}(2010)\citenamefont{Lobo, Dai, Nagel, R\~o\
  om, Carbotte, Timusk, Forget, and Colson}}]{Lobo2010}
\bibinfo{author}{\bibfnamefont{R.~P. S.~M.} \bibnamefont{Lobo}},
  \bibinfo{author}{\bibfnamefont{Y.~M.} \bibnamefont{Dai}},
  \bibinfo{author}{\bibfnamefont{U.}~\bibnamefont{Nagel}},
  \bibinfo{author}{\bibfnamefont{T.}~\bibnamefont{R\~o\ om}},
  \bibinfo{author}{\bibfnamefont{J.~P.} \bibnamefont{Carbotte}},
  \bibinfo{author}{\bibfnamefont{T.}~\bibnamefont{Timusk}},
  \bibinfo{author}{\bibfnamefont{A.}~\bibnamefont{Forget}}, \bibnamefont{and}
  \bibinfo{author}{\bibfnamefont{D.}~\bibnamefont{Colson}},
  \bibinfo{journal}{Phys. Rev. B} \textbf{\bibinfo{volume}{82}},
  \bibinfo{pages}{100506} (\bibinfo{year}{2010}).

\bibitem[{\citenamefont{Tu et~al.}(2010)\citenamefont{Tu, Li, Liu, Punnoose,
  Gong, Ren, Li, Cao, Xu, and Homes}}]{Tu2010}
\bibinfo{author}{\bibfnamefont{J.~J.} \bibnamefont{Tu}},
  \bibinfo{author}{\bibfnamefont{J.}~\bibnamefont{Li}},
  \bibinfo{author}{\bibfnamefont{W.}~\bibnamefont{Liu}},
  \bibinfo{author}{\bibfnamefont{A.}~\bibnamefont{Punnoose}},
  \bibinfo{author}{\bibfnamefont{Y.}~\bibnamefont{Gong}},
  \bibinfo{author}{\bibfnamefont{Y.~H.} \bibnamefont{Ren}},
  \bibinfo{author}{\bibfnamefont{L.~J.} \bibnamefont{Li}},
  \bibinfo{author}{\bibfnamefont{G.~H.} \bibnamefont{Cao}},
  \bibinfo{author}{\bibfnamefont{Z.~A.} \bibnamefont{Xu}}, \bibnamefont{and}
  \bibinfo{author}{\bibfnamefont{C.~C.} \bibnamefont{Homes}},
  \bibinfo{journal}{Phys. Rev. B} \textbf{\bibinfo{volume}{82}},
  \bibinfo{pages}{174509} (\bibinfo{year}{2010}).

\bibitem[{\citenamefont{Teague et~al.}(2011)\citenamefont{Teague, Drayna,
  Lockhart, Cheng, Shen, Wen, and Yeh}}]{Teague2011}
\bibinfo{author}{\bibfnamefont{M.~L.} \bibnamefont{Teague}},
  \bibinfo{author}{\bibfnamefont{G.~K.} \bibnamefont{Drayna}},
  \bibinfo{author}{\bibfnamefont{G.~P.} \bibnamefont{Lockhart}},
  \bibinfo{author}{\bibfnamefont{P.}~\bibnamefont{Cheng}},
  \bibinfo{author}{\bibfnamefont{B.}~\bibnamefont{Shen}},
  \bibinfo{author}{\bibfnamefont{H.-H.} \bibnamefont{Wen}}, \bibnamefont{and}
  \bibinfo{author}{\bibfnamefont{N.-C.} \bibnamefont{Yeh}},
  \bibinfo{journal}{Phys. Rev. Lett.} \textbf{\bibinfo{volume}{106}},
  \bibinfo{pages}{087004} (\bibinfo{year}{2011}).

\bibitem[{\citenamefont{Terashima et~al.}(2009)\citenamefont{Terashima, Sekiba,
  Bowen, Nakayama, Kawahara, Sato, Richard, Xu, Li, Cao
  et~al.}}]{Terashima2009}
\bibinfo{author}{\bibfnamefont{K.}~\bibnamefont{Terashima}},
  \bibinfo{author}{\bibfnamefont{Y.}~\bibnamefont{Sekiba}},
  \bibinfo{author}{\bibfnamefont{J.~H.} \bibnamefont{Bowen}},
  \bibinfo{author}{\bibfnamefont{K.}~\bibnamefont{Nakayama}},
  \bibinfo{author}{\bibfnamefont{T.}~\bibnamefont{Kawahara}},
  \bibinfo{author}{\bibfnamefont{T.}~\bibnamefont{Sato}},
  \bibinfo{author}{\bibfnamefont{P.}~\bibnamefont{Richard}},
  \bibinfo{author}{\bibfnamefont{Y.~M.} \bibnamefont{Xu}},
  \bibinfo{author}{\bibfnamefont{L.~J.} \bibnamefont{Li}},
  \bibinfo{author}{\bibfnamefont{G.~H.} \bibnamefont{Cao}},
  \bibnamefont{et~al.}, \bibinfo{journal}{Proc. Natl. Acad. Sci.}
  \textbf{\bibinfo{volume}{106}}, \bibinfo{pages}{7330} (\bibinfo{year}{2009}).

\bibitem[{\citenamefont{Li et~al.}(2008)\citenamefont{Li, Hu, Dong, Li, Zheng,
  Chen, Luo, and Wang}}]{Li2008}
\bibinfo{author}{\bibfnamefont{G.}~\bibnamefont{Li}},
  \bibinfo{author}{\bibfnamefont{W.~Z.} \bibnamefont{Hu}},
  \bibinfo{author}{\bibfnamefont{J.}~\bibnamefont{Dong}},
  \bibinfo{author}{\bibfnamefont{Z.}~\bibnamefont{Li}},
  \bibinfo{author}{\bibfnamefont{P.}~\bibnamefont{Zheng}},
  \bibinfo{author}{\bibfnamefont{G.~F.} \bibnamefont{Chen}},
  \bibinfo{author}{\bibfnamefont{J.~L.} \bibnamefont{Luo}}, \bibnamefont{and}
  \bibinfo{author}{\bibfnamefont{N.~L.} \bibnamefont{Wang}},
  \bibinfo{journal}{Phys. Rev. Lett.} \textbf{\bibinfo{volume}{101}},
  \bibinfo{pages}{107004} (\bibinfo{year}{2008}).

\bibitem[{\citenamefont{Charnukha et~al.}(2011)\citenamefont{Charnukha, Dolgov,
  Golubov, Matiks, Sun, Lin, Keimer, and Boris}}]{Charnukha2011a}
\bibinfo{author}{\bibfnamefont{A.}~\bibnamefont{Charnukha}},
  \bibinfo{author}{\bibfnamefont{O.~V.} \bibnamefont{Dolgov}},
  \bibinfo{author}{\bibfnamefont{A.~A.} \bibnamefont{Golubov}},
  \bibinfo{author}{\bibfnamefont{Y.}~\bibnamefont{Matiks}},
  \bibinfo{author}{\bibfnamefont{D.~L.} \bibnamefont{Sun}},
  \bibinfo{author}{\bibfnamefont{C.~T.} \bibnamefont{Lin}},
  \bibinfo{author}{\bibfnamefont{B.}~\bibnamefont{Keimer}}, \bibnamefont{and}
  \bibinfo{author}{\bibfnamefont{A.~V.} \bibnamefont{Boris}},
  \bibinfo{journal}{Phys. Rev. B} \textbf{\bibinfo{volume}{84}},
  \bibinfo{pages}{174511} (\bibinfo{year}{2011}).

\bibitem[{\citenamefont{Dai et~al.}(2011)\citenamefont{Dai, Xu, Shen, Wen, Qiu,
  and Lobo}}]{Dai2011}
\bibinfo{author}{\bibfnamefont{Y.~M.} \bibnamefont{Dai}},
  \bibinfo{author}{\bibfnamefont{B.}~\bibnamefont{Xu}},
  \bibinfo{author}{\bibfnamefont{B.}~\bibnamefont{Shen}},
  \bibinfo{author}{\bibfnamefont{H.~H.} \bibnamefont{Wen}},
  \bibinfo{author}{\bibfnamefont{X.~G.} \bibnamefont{Qiu}}, \bibnamefont{and}
  \bibinfo{author}{\bibfnamefont{R.~P. S.~M.} \bibnamefont{Lobo}},
  \bibinfo{journal}{arXiv} \textbf{\bibinfo{volume}{1106}},
  \bibinfo{pages}{4430v1} (\bibinfo{year}{2011}).

\bibitem[{\citenamefont{Shan et~al.}(2011)\citenamefont{Shan, Wang, Gong, Shen,
  Huang, Yang, Ren, and Wen}}]{Shan2011}
\bibinfo{author}{\bibfnamefont{L.}~\bibnamefont{Shan}},
  \bibinfo{author}{\bibfnamefont{Y.-L.} \bibnamefont{Wang}},
  \bibinfo{author}{\bibfnamefont{J.}~\bibnamefont{Gong}},
  \bibinfo{author}{\bibfnamefont{B.}~\bibnamefont{Shen}},
  \bibinfo{author}{\bibfnamefont{Y.}~\bibnamefont{Huang}},
  \bibinfo{author}{\bibfnamefont{H.}~\bibnamefont{Yang}},
  \bibinfo{author}{\bibfnamefont{C.}~\bibnamefont{Ren}}, \bibnamefont{and}
  \bibinfo{author}{\bibfnamefont{H.-H.} \bibnamefont{Wen}},
  \bibinfo{journal}{Phys. Rev. B} \textbf{\bibinfo{volume}{83}},
  \bibinfo{pages}{060510} (\bibinfo{year}{2011}).

\bibitem[{\citenamefont{Ding et~al.}(2008)\citenamefont{Ding, Richard,
  Nakayama, Sugawara, Arakane, Sekiba, Takayama, Souma, Sato, Takahashi
  et~al.}}]{Ding2008}
\bibinfo{author}{\bibfnamefont{H.}~\bibnamefont{Ding}},
  \bibinfo{author}{\bibfnamefont{P.}~\bibnamefont{Richard}},
  \bibinfo{author}{\bibfnamefont{K.}~\bibnamefont{Nakayama}},
  \bibinfo{author}{\bibfnamefont{K.}~\bibnamefont{Sugawara}},
  \bibinfo{author}{\bibfnamefont{T.}~\bibnamefont{Arakane}},
  \bibinfo{author}{\bibfnamefont{Y.}~\bibnamefont{Sekiba}},
  \bibinfo{author}{\bibfnamefont{A.}~\bibnamefont{Takayama}},
  \bibinfo{author}{\bibfnamefont{S.}~\bibnamefont{Souma}},
  \bibinfo{author}{\bibfnamefont{T.}~\bibnamefont{Sato}},
  \bibinfo{author}{\bibfnamefont{T.}~\bibnamefont{Takahashi}},
  \bibnamefont{et~al.}, \bibinfo{journal}{EPL (Europhysics Letters)}
  \textbf{\bibinfo{volume}{83}}, \bibinfo{pages}{47001} (\bibinfo{year}{2008}).

\bibitem[{\citenamefont{Park et~al.}(2009)\citenamefont{Park, Inosov,
  Niedermayer, Sun, Haug, Christensen, Dinnebier, Boris, Drew, Schulz
  et~al.}}]{Park2009}
\bibinfo{author}{\bibfnamefont{J.~T.} \bibnamefont{Park}},
  \bibinfo{author}{\bibfnamefont{D.~S.} \bibnamefont{Inosov}},
  \bibinfo{author}{\bibfnamefont{C.}~\bibnamefont{Niedermayer}},
  \bibinfo{author}{\bibfnamefont{G.~L.} \bibnamefont{Sun}},
  \bibinfo{author}{\bibfnamefont{D.}~\bibnamefont{Haug}},
  \bibinfo{author}{\bibfnamefont{N.~B.} \bibnamefont{Christensen}},
  \bibinfo{author}{\bibfnamefont{R.}~\bibnamefont{Dinnebier}},
  \bibinfo{author}{\bibfnamefont{A.~V.} \bibnamefont{Boris}},
  \bibinfo{author}{\bibfnamefont{A.~J.} \bibnamefont{Drew}},
  \bibinfo{author}{\bibfnamefont{L.}~\bibnamefont{Schulz}},
  \bibnamefont{et~al.}, \bibinfo{journal}{Phys. Rev. Lett.}
  \textbf{\bibinfo{volume}{102}}, \bibinfo{pages}{117006}
  (\bibinfo{year}{2009}).

\bibitem[{\citenamefont{Goko et~al.}(2009)\citenamefont{Goko, Aczel,
  Baggio-Saitovitch, Bud'ko, Canfield, Carlo, Chen, Dai, Hamann, Hu
  et~al.}}]{Goko2009}
\bibinfo{author}{\bibfnamefont{T.}~\bibnamefont{Goko}},
  \bibinfo{author}{\bibfnamefont{A.~A.} \bibnamefont{Aczel}},
  \bibinfo{author}{\bibfnamefont{E.}~\bibnamefont{Baggio-Saitovitch}},
  \bibinfo{author}{\bibfnamefont{S.~L.} \bibnamefont{Bud'ko}},
  \bibinfo{author}{\bibfnamefont{P.~C.} \bibnamefont{Canfield}},
  \bibinfo{author}{\bibfnamefont{J.~P.} \bibnamefont{Carlo}},
  \bibinfo{author}{\bibfnamefont{G.~F.} \bibnamefont{Chen}},
  \bibinfo{author}{\bibfnamefont{P.}~\bibnamefont{Dai}},
  \bibinfo{author}{\bibfnamefont{A.~C.} \bibnamefont{Hamann}},
  \bibinfo{author}{\bibfnamefont{W.~Z.} \bibnamefont{Hu}},
  \bibnamefont{et~al.}, \bibinfo{journal}{Phys. Rev. B}
  \textbf{\bibinfo{volume}{80}}, \bibinfo{pages}{024508}
  (\bibinfo{year}{2009}).

\bibitem[{\citenamefont{Aczel et~al.}(2008)\citenamefont{Aczel,
  Baggio-Saitovitch, Budko, Canfield, Carlo, Chen, Dai, Goko, Hu, Luke
  et~al.}}]{Aczel2008}
\bibinfo{author}{\bibfnamefont{A.~A.} \bibnamefont{Aczel}},
  \bibinfo{author}{\bibfnamefont{E.}~\bibnamefont{Baggio-Saitovitch}},
  \bibinfo{author}{\bibfnamefont{S.~L.} \bibnamefont{Budko}},
  \bibinfo{author}{\bibfnamefont{P.~C.} \bibnamefont{Canfield}},
  \bibinfo{author}{\bibfnamefont{J.~P.} \bibnamefont{Carlo}},
  \bibinfo{author}{\bibfnamefont{G.~F.} \bibnamefont{Chen}},
  \bibinfo{author}{\bibfnamefont{P.}~\bibnamefont{Dai}},
  \bibinfo{author}{\bibfnamefont{T.}~\bibnamefont{Goko}},
  \bibinfo{author}{\bibfnamefont{W.~Z.} \bibnamefont{Hu}},
  \bibinfo{author}{\bibfnamefont{G.~M.} \bibnamefont{Luke}},
  \bibnamefont{et~al.}, \bibinfo{journal}{Phys. Rev. B}
  \textbf{\bibinfo{volume}{78}}, \bibinfo{pages}{214503}
  (\bibinfo{year}{2008}).

\bibitem[{\citenamefont{Massee et~al.}(2009)\citenamefont{Massee, Huang,
  Huisman, de~Jong, Goedkoop, and Golden}}]{Massee2009}
\bibinfo{author}{\bibfnamefont{F.}~\bibnamefont{Massee}},
  \bibinfo{author}{\bibfnamefont{Y.}~\bibnamefont{Huang}},
  \bibinfo{author}{\bibfnamefont{R.}~\bibnamefont{Huisman}},
  \bibinfo{author}{\bibfnamefont{S.}~\bibnamefont{de~Jong}},
  \bibinfo{author}{\bibfnamefont{J.~B.} \bibnamefont{Goedkoop}},
  \bibnamefont{and} \bibinfo{author}{\bibfnamefont{M.~S.}
  \bibnamefont{Golden}}, \bibinfo{journal}{Phys. Rev. B}
  \textbf{\bibinfo{volume}{79}}, \bibinfo{pages}{220517}
  (\bibinfo{year}{2009}).

\bibitem[{\citenamefont{Chia et~al.}(2010)\citenamefont{Chia, Talbayev, Zhu,
  Yuan, Park, Thompson, Panagopoulos, Chen, Luo, Wang et~al.}}]{Chia2010}
\bibinfo{author}{\bibfnamefont{E.~E.~M.} \bibnamefont{Chia}},
  \bibinfo{author}{\bibfnamefont{D.}~\bibnamefont{Talbayev}},
  \bibinfo{author}{\bibfnamefont{J.-X.} \bibnamefont{Zhu}},
  \bibinfo{author}{\bibfnamefont{H.~Q.} \bibnamefont{Yuan}},
  \bibinfo{author}{\bibfnamefont{T.}~\bibnamefont{Park}},
  \bibinfo{author}{\bibfnamefont{J.~D.} \bibnamefont{Thompson}},
  \bibinfo{author}{\bibfnamefont{C.}~\bibnamefont{Panagopoulos}},
  \bibinfo{author}{\bibfnamefont{G.~F.} \bibnamefont{Chen}},
  \bibinfo{author}{\bibfnamefont{J.~L.} \bibnamefont{Luo}},
  \bibinfo{author}{\bibfnamefont{N.~L.} \bibnamefont{Wang}},
  \bibnamefont{et~al.}, \bibinfo{journal}{Phys. Rev. Lett.}
  \textbf{\bibinfo{volume}{104}}, \bibinfo{pages}{027003}
  (\bibinfo{year}{2010}).

\bibitem[{\citenamefont{Xu et~al.}(2011)\citenamefont{Xu, Richard, Nakayama,
  Kawahara, Sekiba, Qian, Neupane, Souma, Sato, Takahashi et~al.}}]{Xu2011}
\bibinfo{author}{\bibfnamefont{Y.-M.} \bibnamefont{Xu}},
  \bibinfo{author}{\bibfnamefont{P.}~\bibnamefont{Richard}},
  \bibinfo{author}{\bibfnamefont{K.}~\bibnamefont{Nakayama}},
  \bibinfo{author}{\bibfnamefont{T.}~\bibnamefont{Kawahara}},
  \bibinfo{author}{\bibfnamefont{Y.}~\bibnamefont{Sekiba}},
  \bibinfo{author}{\bibfnamefont{T.}~\bibnamefont{Qian}},
  \bibinfo{author}{\bibfnamefont{M.}~\bibnamefont{Neupane}},
  \bibinfo{author}{\bibfnamefont{S.}~\bibnamefont{Souma}},
  \bibinfo{author}{\bibfnamefont{T.}~\bibnamefont{Sato}},
  \bibinfo{author}{\bibfnamefont{T.}~\bibnamefont{Takahashi}},
  \bibnamefont{et~al.}, \bibinfo{journal}{Nat Commun}
  \textbf{\bibinfo{volume}{2}}, \bibinfo{pages}{392} (\bibinfo{year}{2011}).

\bibitem[{\citenamefont{Yu et~al.}(2008)\citenamefont{Yu, Munzar, Boris,
  Yordanov, Chaloupka, Wolf, Lin, Keimer, and Bernhard}}]{Yu2008}
\bibinfo{author}{\bibfnamefont{L.}~\bibnamefont{Yu}},
  \bibinfo{author}{\bibfnamefont{D.}~\bibnamefont{Munzar}},
  \bibinfo{author}{\bibfnamefont{A.~V.} \bibnamefont{Boris}},
  \bibinfo{author}{\bibfnamefont{P.}~\bibnamefont{Yordanov}},
  \bibinfo{author}{\bibfnamefont{J.}~\bibnamefont{Chaloupka}},
  \bibinfo{author}{\bibfnamefont{T.}~\bibnamefont{Wolf}},
  \bibinfo{author}{\bibfnamefont{C.~T.} \bibnamefont{Lin}},
  \bibinfo{author}{\bibfnamefont{B.}~\bibnamefont{Keimer}}, \bibnamefont{and}
  \bibinfo{author}{\bibfnamefont{C.}~\bibnamefont{Bernhard}},
  \bibinfo{journal}{Phys. Rev. Lett.} \textbf{\bibinfo{volume}{100}},
  \bibinfo{pages}{177004} (\bibinfo{year}{2008}).

\bibitem[{\citenamefont{Luo et~al.}(2008)\citenamefont{Luo, Wang, Yang, Cheng,
  Zhu, and Wen}}]{Luo2008}
\bibinfo{author}{\bibfnamefont{H.}~\bibnamefont{Luo}},
  \bibinfo{author}{\bibfnamefont{Z.}~\bibnamefont{Wang}},
  \bibinfo{author}{\bibfnamefont{H.}~\bibnamefont{Yang}},
  \bibinfo{author}{\bibfnamefont{P.}~\bibnamefont{Cheng}},
  \bibinfo{author}{\bibfnamefont{X.}~\bibnamefont{Zhu}}, \bibnamefont{and}
  \bibinfo{author}{\bibfnamefont{H.-H.} \bibnamefont{Wen}},
  \bibinfo{journal}{Superconductor Science and Technology}
  \textbf{\bibinfo{volume}{21}}, \bibinfo{pages}{125014}
  (\bibinfo{year}{2008}).

\bibitem[{\citenamefont{Homes et~al.}(1993)\citenamefont{Homes, Timusk, Liang,
  Bonn, and Hardy}}]{Homes1993}
\bibinfo{author}{\bibfnamefont{C.~C.} \bibnamefont{Homes}},
  \bibinfo{author}{\bibfnamefont{T.}~\bibnamefont{Timusk}},
  \bibinfo{author}{\bibfnamefont{R.}~\bibnamefont{Liang}},
  \bibinfo{author}{\bibfnamefont{D.~A.} \bibnamefont{Bonn}}, \bibnamefont{and}
  \bibinfo{author}{\bibfnamefont{W.~N.} \bibnamefont{Hardy}},
  \bibinfo{journal}{Phys. Rev. Lett.} \textbf{\bibinfo{volume}{71}},
  \bibinfo{pages}{1645} (\bibinfo{year}{1993}).

\bibitem[{\citenamefont{Nakajima et~al.}(2010)\citenamefont{Nakajima, Ishida,
  Kihou, Tomioka, Ito, Yoshida, Lee, Kito, Iyo, Eisaki et~al.}}]{Nakajima2010}
\bibinfo{author}{\bibfnamefont{M.}~\bibnamefont{Nakajima}},
  \bibinfo{author}{\bibfnamefont{S.}~\bibnamefont{Ishida}},
  \bibinfo{author}{\bibfnamefont{K.}~\bibnamefont{Kihou}},
  \bibinfo{author}{\bibfnamefont{Y.}~\bibnamefont{Tomioka}},
  \bibinfo{author}{\bibfnamefont{T.}~\bibnamefont{Ito}},
  \bibinfo{author}{\bibfnamefont{Y.}~\bibnamefont{Yoshida}},
  \bibinfo{author}{\bibfnamefont{C.~H.} \bibnamefont{Lee}},
  \bibinfo{author}{\bibfnamefont{H.}~\bibnamefont{Kito}},
  \bibinfo{author}{\bibfnamefont{A.}~\bibnamefont{Iyo}},
  \bibinfo{author}{\bibfnamefont{H.}~\bibnamefont{Eisaki}},
  \bibnamefont{et~al.}, \bibinfo{journal}{Phys. Rev. B}
  \textbf{\bibinfo{volume}{81}}, \bibinfo{pages}{104528}
  (\bibinfo{year}{2010}).

\bibitem[{\citenamefont{Pfuner et~al.}(2009)\citenamefont{Pfuner, Analytis,
  Chu, Fisher, and Degiorgi}}]{Pfuner2009}
\bibinfo{author}{\bibfnamefont{F.}~\bibnamefont{Pfuner}},
  \bibinfo{author}{\bibfnamefont{J.~G.} \bibnamefont{Analytis}},
  \bibinfo{author}{\bibfnamefont{J.-H.} \bibnamefont{Chu}},
  \bibinfo{author}{\bibfnamefont{I.~R.} \bibnamefont{Fisher}},
  \bibnamefont{and} \bibinfo{author}{\bibfnamefont{L.}~\bibnamefont{Degiorgi}},
  \bibinfo{journal}{The European Physical Journal B - Condensed Matter and
  Complex Systems} \textbf{\bibinfo{volume}{67}}, \bibinfo{pages}{513}
  (\bibinfo{year}{2009}), ISSN \bibinfo{issn}{1434-6028}.

\bibitem[{\citenamefont{Zimmers et~al.}(2004)\citenamefont{Zimmers, Lobo,
  Bontemps, Homes, Barr, Dagan, and Greene}}]{Zimmers2004}
\bibinfo{author}{\bibfnamefont{A.}~\bibnamefont{Zimmers}},
  \bibinfo{author}{\bibfnamefont{R.~P. S.~M.} \bibnamefont{Lobo}},
  \bibinfo{author}{\bibfnamefont{N.}~\bibnamefont{Bontemps}},
  \bibinfo{author}{\bibfnamefont{C.~C.} \bibnamefont{Homes}},
  \bibinfo{author}{\bibfnamefont{M.~C.} \bibnamefont{Barr}},
  \bibinfo{author}{\bibfnamefont{Y.}~\bibnamefont{Dagan}}, \bibnamefont{and}
  \bibinfo{author}{\bibfnamefont{R.~L.} \bibnamefont{Greene}},
  \bibinfo{journal}{Phys. Rev. B} \textbf{\bibinfo{volume}{70}},
  \bibinfo{pages}{132502} (\bibinfo{year}{2004}).

\bibitem[{\citenamefont{Dordevic et~al.}(2002)\citenamefont{Dordevic, Singley,
  Basov, Komiya, Ando, Bucher, Homes, and Strongin}}]{Dordevic2002}
\bibinfo{author}{\bibfnamefont{S.~V.} \bibnamefont{Dordevic}},
  \bibinfo{author}{\bibfnamefont{E.~J.} \bibnamefont{Singley}},
  \bibinfo{author}{\bibfnamefont{D.~N.} \bibnamefont{Basov}},
  \bibinfo{author}{\bibfnamefont{S.}~\bibnamefont{Komiya}},
  \bibinfo{author}{\bibfnamefont{Y.}~\bibnamefont{Ando}},
  \bibinfo{author}{\bibfnamefont{E.}~\bibnamefont{Bucher}},
  \bibinfo{author}{\bibfnamefont{C.~C.} \bibnamefont{Homes}}, \bibnamefont{and}
  \bibinfo{author}{\bibfnamefont{M.}~\bibnamefont{Strongin}},
  \bibinfo{journal}{Phys. Rev. B} \textbf{\bibinfo{volume}{65}},
  \bibinfo{pages}{134511} (\bibinfo{year}{2002}).

\bibitem[{\citenamefont{Julien et~al.}(2009)\citenamefont{Julien, Mayaffre,
  Horvati\'{c}, Berthier, Zhang, Wu, Chen, Wang, and Luo}}]{Julien2009}
\bibinfo{author}{\bibfnamefont{M.-H.} \bibnamefont{Julien}},
  \bibinfo{author}{\bibfnamefont{H.}~\bibnamefont{Mayaffre}},
  \bibinfo{author}{\bibfnamefont{M.}~\bibnamefont{Horvati\'{c}}},
  \bibinfo{author}{\bibfnamefont{C.}~\bibnamefont{Berthier}},
  \bibinfo{author}{\bibfnamefont{X.~D.} \bibnamefont{Zhang}},
  \bibinfo{author}{\bibfnamefont{W.}~\bibnamefont{Wu}},
  \bibinfo{author}{\bibfnamefont{G.~F.} \bibnamefont{Chen}},
  \bibinfo{author}{\bibfnamefont{N.~L.} \bibnamefont{Wang}}, \bibnamefont{and}
  \bibinfo{author}{\bibfnamefont{J.~L.} \bibnamefont{Luo}},
  \bibinfo{journal}{EPL (Europhysics Letters)} \textbf{\bibinfo{volume}{87}},
  \bibinfo{pages}{37001} (\bibinfo{year}{2009}).

\bibitem[{\citenamefont{Reid et~al.}(2011)\citenamefont{Reid, Tanatar, Luo,
  Shakeripour, de~Cotret, Doiron-Leyraud, Chang, Shen, Wen, Kim
  et~al.}}]{Reid2011}
\bibinfo{author}{\bibfnamefont{J.-P.} \bibnamefont{Reid}},
  \bibinfo{author}{\bibfnamefont{M.~A.} \bibnamefont{Tanatar}},
  \bibinfo{author}{\bibfnamefont{X.~G.} \bibnamefont{Luo}},
  \bibinfo{author}{\bibfnamefont{H.}~\bibnamefont{Shakeripour}},
  \bibinfo{author}{\bibfnamefont{S.~R.} \bibnamefont{de~Cotret}},
  \bibinfo{author}{\bibfnamefont{N.}~\bibnamefont{Doiron-Leyraud}},
  \bibinfo{author}{\bibfnamefont{J.}~\bibnamefont{Chang}},
  \bibinfo{author}{\bibfnamefont{B.}~\bibnamefont{Shen}},
  \bibinfo{author}{\bibfnamefont{H.-H.} \bibnamefont{Wen}},
  \bibinfo{author}{\bibfnamefont{H.}~\bibnamefont{Kim}}, \bibnamefont{et~al.},
  \bibinfo{journal}{arXiv} \textbf{\bibinfo{volume}{1105}},
  \bibinfo{pages}{2232v1} (\bibinfo{year}{2011}).

\bibitem[{\citenamefont{Maiti et~al.}(2012)\citenamefont{Maiti, Fernandes, and
  Chubukov}}]{Maiti2012}
\bibinfo{author}{\bibfnamefont{S.}~\bibnamefont{Maiti}},
  \bibinfo{author}{\bibfnamefont{R.~M.} \bibnamefont{Fernandes}},
  \bibnamefont{and} \bibinfo{author}{\bibfnamefont{A.~V.}
  \bibnamefont{Chubukov}}, \bibinfo{journal}{arXiv}
  \textbf{\bibinfo{volume}{1203}}, \bibinfo{pages}{0991}
  (\bibinfo{year}{2012}).

\bibitem[{\citenamefont{Ioffe and Millis}(1999)}]{Ioffe1999}
\bibinfo{author}{\bibfnamefont{L.~B.} \bibnamefont{Ioffe}} \bibnamefont{and}
  \bibinfo{author}{\bibfnamefont{A.~J.} \bibnamefont{Millis}},
  \bibinfo{journal}{Science} \textbf{\bibinfo{volume}{285}},
  \bibinfo{pages}{1241} (\bibinfo{year}{1999}).

\bibitem[{\citenamefont{Ioffe and Millis}(2000)}]{Ioffe2000}
\bibinfo{author}{\bibfnamefont{L.~B.} \bibnamefont{Ioffe}} \bibnamefont{and}
  \bibinfo{author}{\bibfnamefont{A.~J.} \bibnamefont{Millis}},
  \bibinfo{journal}{Phys. Rev. B} \textbf{\bibinfo{volume}{61}},
  \bibinfo{pages}{9077} (\bibinfo{year}{2000}).

\bibitem[{\citenamefont{Baek et~al.}(2011)\citenamefont{Baek, Grafe, Harnagea,
  Singh, Wurmehl, and B\"uchner}}]{Baek2011}
\bibinfo{author}{\bibfnamefont{S.-H.} \bibnamefont{Baek}},
  \bibinfo{author}{\bibfnamefont{H.-J.} \bibnamefont{Grafe}},
  \bibinfo{author}{\bibfnamefont{L.}~\bibnamefont{Harnagea}},
  \bibinfo{author}{\bibfnamefont{S.}~\bibnamefont{Singh}},
  \bibinfo{author}{\bibfnamefont{S.}~\bibnamefont{Wurmehl}}, \bibnamefont{and}
  \bibinfo{author}{\bibfnamefont{B.}~\bibnamefont{B\"uchner}},
  \bibinfo{journal}{Phys. Rev. B} \textbf{\bibinfo{volume}{84}},
  \bibinfo{pages}{094510} (\bibinfo{year}{2011}).

\bibitem[{\citenamefont{Sheet et~al.}(2010)\citenamefont{Sheet, Mehta, Dikin,
  Lee, Bark, Jiang, Weiss, Hellstrom, Rzchowski, Eom et~al.}}]{Sheet2010}
\bibinfo{author}{\bibfnamefont{G.}~\bibnamefont{Sheet}},
  \bibinfo{author}{\bibfnamefont{M.}~\bibnamefont{Mehta}},
  \bibinfo{author}{\bibfnamefont{D.~A.} \bibnamefont{Dikin}},
  \bibinfo{author}{\bibfnamefont{S.}~\bibnamefont{Lee}},
  \bibinfo{author}{\bibfnamefont{C.~W.} \bibnamefont{Bark}},
  \bibinfo{author}{\bibfnamefont{J.}~\bibnamefont{Jiang}},
  \bibinfo{author}{\bibfnamefont{J.~D.} \bibnamefont{Weiss}},
  \bibinfo{author}{\bibfnamefont{E.~E.} \bibnamefont{Hellstrom}},
  \bibinfo{author}{\bibfnamefont{M.~S.} \bibnamefont{Rzchowski}},
  \bibinfo{author}{\bibfnamefont{C.~B.} \bibnamefont{Eom}},
  \bibnamefont{et~al.}, \bibinfo{journal}{Phys. Rev. Lett.}
  \textbf{\bibinfo{volume}{105}}, \bibinfo{pages}{167003}
  (\bibinfo{year}{2010}).

\end{thebibliography}
\end{document}